\documentclass{egpubl}
\usepackage{eurovis2023}

\SpecialIssuePaper         

\CGFStandardLicense

\usepackage[T1]{fontenc}
\usepackage{dfadobe}  

\usepackage{cite}  
\BibtexOrBiblatex
\electronicVersion
\PrintedOrElectronic
\ifpdf \usepackage[pdftex]{graphicx} \pdfcompresslevel=9
\else \usepackage[dvips]{graphicx} \fi

\usepackage{egweblnk}
\usepackage{graphicx}
\usepackage{amsmath}
\usepackage{amssymb}
\usepackage{booktabs}
\usepackage{bm}
\usepackage{caption}
\usepackage{float}
\usepackage{array}
\usepackage{tabularx}
\usepackage{booktabs}
\usepackage{ragged2e}
\usepackage{color, colortbl}
\usepackage{makecell}
\usepackage{caption}
\usepackage{bbding}
\usepackage{multirow}
\usepackage{fontawesome}
\usepackage{utfsym}
\usepackage{colortbl}
\usepackage{multicol}

\definecolor{Gray}{gray}{0.9}
\definecolor{LightCyan}{rgb}{0.88,1,1}
\definecolor{green}{rgb}{0,0.41,0.09}
\definecolor{blue}{rgb}{0,0,0.7}
\definecolor{prompt}{rgb}{0.5, 0.5, 0.5}
\definecolor{extended}{rgb}{0.33, 0.8, 0.725}
\definecolor{primary}{rgb}{0,0.2, 0.572}
\newcommand{\eg}{\emph{e.g.}}

\newcommand{\primary}[1]{{\color{primary}{#1}}}
\newcommand{\extended}[1]{{\color{extended}{#1}}}
\newcommand{\prompt}[1]{{\color{prompt}{\textours{#1}}}}

\newcommand{\revise}[1]{{\color{black}{#1}}}
\DeclareTextFontCommand{\textours}{\fontfamily{qpl}\selectfont}

\title[WYTIWYR: A User Intent-Aware Framework with Multi-modal Inputs for Visualization Retrieval]
{WYTIWYR: A User Intent-Aware Framework with Multi-modal Inputs for Visualization Retrieval}

\author[S. Xiao et al.]
{\parbox{\textwidth}{\centering Shishi Xiao$^{1}$\orcid{0009-0008-0262-5289}, Yihan Hou$^{1}$\orcid{0000-0002-1459-8766}, Cheng Jin$^{2}$\orcid{0000-0002-3522-3592}, and Wei Zeng$^{1,2}$\thanks{Wei Zeng is the corresponding author. Email: weizeng@ust.hk}\orcid{0000-0002-5600-8824}
        }
        \\
{\parbox{\textwidth}{\centering $^1$The Hong Kong University of Science and Technology (Guangzhou), Guangzhou, China\\
         $^2$ The Hong Kong University of Science and Technology, Hong Kong SAR, China
      }
}
}

\begin{document}

\maketitle
\begin{abstract}
Retrieving charts from a large corpus is a fundamental task that can benefit numerous applications such as visualization recommendations.
The retrieved results are expected to conform to both explicit visual attributes (\eg, chart type, colormap) and implicit user intents (\eg, design style, context information) that vary upon application scenarios.
However, existing example-based chart retrieval methods are built upon non-decoupled and low-level visual features that are hard to interpret, while definition-based ones are constrained to pre-defined attributes that are hard to extend.
In this work, we propose a new framework, namely \emph{WYTIWYR (What-You-Think-Is-What-You-Retrieve)}, that integrates user intents into the chart retrieval process.
The framework consists of two stages: first, the \emph{Annotation} stage disentangles the visual attributes within the query chart; and second, the \emph{Retrieval} stage embeds the user's intent with customized text prompt as well as bitmap query chart, to recall targeted retrieval result. 
We develop a prototype \emph{WYTIWYR} system leveraging a contrastive language-image pre-training (CLIP) model to achieve zero-shot classification as well as multi-modal input encoding, and test the prototype on a large corpus with charts crawled from the Internet.
Quantitative experiments, case studies, and qualitative interviews are conducted.
The results demonstrate the usability and effectiveness of our proposed framework.


\end{abstract}  
\section{Introduction}
\label{sec: intro}
Data visualization can empower users' understanding of data and facilitate communication. Therefore, enormous charts are flourishing on the Internet.
Designing an appropriate chart is a time-consuming and labor-intensive process that needs to consider various visual attributes (\eg, chart type, color) ~\cite{shneiderman_1996_eyes}, as well as the design style (\eg, context information, 2D/3D effect) ~\cite{moere2012evaluating}.
Therefore, designing based on existing examples, rather than starting with sketches, is a more preferred approach ~\cite{battle2021exploring, parsons2021fixation}.
Chart retrieval, as an approach to return the ranked examples with respect to the corresponding query, has gained much interest in both industry and academia~\cite{li2022structure, hoque2019searching, siegel2016figureseer}.

\begin{figure}[t]
\centering
\includegraphics[width=0.995\columnwidth]{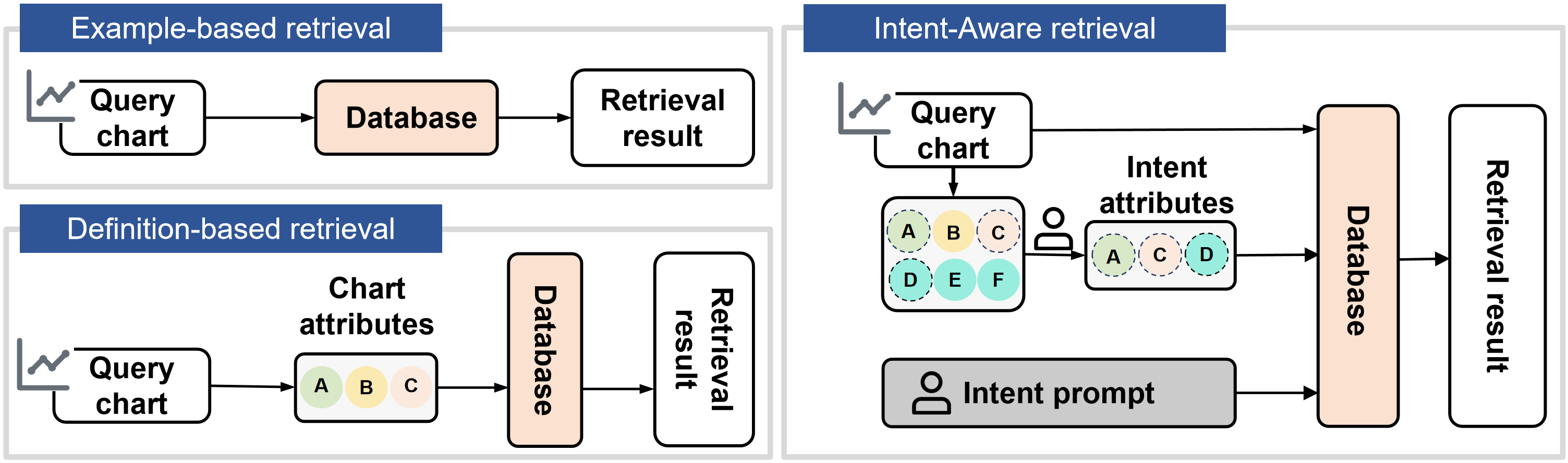} 
\vspace{-5mm}
\caption{Comparison of pipelines of existing and our proposed chart retrieval frameworks: example-based retrieval (left-top), definition-based retrieval (left-bottom), and our proposed intent-aware retrieval (right).}
\vspace{-6mm}
\label{fig1: intro}
\end{figure}

However, existing works mainly focus on improving the similarity between query charts and retrieval results while neglecting implicit user intent. 
As illustrated in Figure~\ref{fig1: intro}, existing chart retrieval frameworks might be divided into two categories: example-based (Figure~\ref{fig1: intro} (left-top)) and definition-based (Figure~\ref{fig1: intro} (left-bottom)) approaches.
Example-based methods (\eg,~\cite{Schneidewind_2006_pixnostics,dang2014scagexplorer, behrisch_2018_quality}) characterize the relevance criteria by the visual similarity of charts, whilst definition-based methods (\eg,~\cite{chen2015diagramflyer, hoque2019searching}) measure similarity based on attributes extracted from the input charts.
However, these methods may suffer from both input and target shifts problems.
First, many existing works take only specific formats (\eg, scalable vector graphics (SVG)~\cite{hoque2019searching, li2022structure}) of visualization as inputs, which are often synthesized with monotonous data distributions and limited chart types that incurs overfitting problem of models embedded in the framework. 
In contrast, most online charts are generally in bitmap formats, and they are highly diverse in terms of chart types and visual styles\cite{battle2018beagle}.
In addition, online charts contain noisy and redundant information, putting higher demands on generalizability and robustness of the retrieval framework.
Second, example-based methods would indiscriminately recall pixel-wise visual similarity in the input chart, while definition-based methods consider only a small number of pre-defined and fixed chart attributes.
Such rigid similarity measurements would return unwanted target charts. Furthermore, it may tend to limited the retrieval to a specific chart attribute, a specific combination of chart attributes, or even chart attributes that are not necessarily enclosed by the query chart.

To mitigate such shifts, we propose a novel chart retrieval framework named \textit{WYTIWYR}. It explicitly extracts visual attributes for the query chart, and offers a set of target charts through a customized retrieval procedure that considers flexible combinations of the visual attributes and users' intents.
As Figure~\ref{fig1: intro} (right) illustrates, our method tears down the query chart flexibly with customizable attributes and actively injects user intent as auxiliary inputs to seek better chart retrieval outputs.

Our intent-aware framework consists of two essential stages, 
the first stage is \textit{Annotation}, which highlights explicitly disentangling the visual attributes for query charts. The disentangled attributes would enable the flexibly for users to combine the attributes representing their intent.
Because these attributes differ amongst chart visualizations, we performed preliminary research with 18 visualization categories to establish the branching of each attribute and what users would anticipate when retrieving charts.
Based on the study, we utilize deep neural networks to train several independent attribute classifiers tailored to four primary visual attributes, namely \{\textit{Type}, \textit{Trend}, \textit{Color}, \textit{Layout}\}.
To enable the resilience to user intent, we leverage the state-of-the-art contrastive language-image pretraining (CLIP)~\cite{radford2021learning} model famous for zero-shot classification. The model allows users for their own creation of the attribute classifier to identify the extended attributes. These attributes are excluded from our preliminary set of classifiers but are still within the query process.

The second stage is \textit{Retrieval}, which highlights the role of human-in-the-loop through intent attributes as prompts.
The intent attributes would condition intent-aware filter, narrow the search scope from a huge collection of charts, and finally generate multiple candidate charts.
The multi-modal encoder fuses context information from both the query chart and user intent prompt into a joint representation.
After that, similarity modeling is conducted between such joint representation and candidate charts.
Ultimately, we rank the similarity scores in decreasing order and output the top-$K$ results as the target charts. 
In summary, our main contributions are summarized as follows:
\begin{itemize}
    \item \textbf{Intent-aware Retrieval.} We propose a novel framework integrating user intent into chart retrieval process. The \textit{Annotation} stage disentangles attributes and enables a flexible combination of attributes. The \textit{Retrieval} stage digests both query chart and user intent as multi-modal inputs to get target retrieval results.
    \item \textbf{Prototype Construction.} We implement a prototype system \revise{combining CLIP model with visual interface} to support intent-aware retrieval. 
    \revise{Dataset, code, pretrained model are released at} \url{https://github.com/SerendipitysX/WYTIWYR}.
    \item \textbf{Extensive Evaluation.} We conduct extensive quantitative experiments, case studies as well as expert interviews to validate the effectiveness of our approach. 
\end{itemize}

\section{Related Works}
\textbf{Chart Attributes}.
Visualization is the process of mapping data to images.
The data can be encoded in different ways, yielding various types of chart attributes that can be categorized in several levels.
Low-level visual stimuli of charts include color, position, and shape~\cite{rodrigues_2006_reviewing}, while high-level taxonomy includes consideration of the object of study, data, design model, and user model~\cite{tory_2004_rethinking}.
The importance of choosing appropriate chart attributes has been highlighted.
Recent studies (\eg,~\cite{majooni2018eye, kim2021automated, li2015novel, shi2022supporting}) stress the effects of chart attributes on users' comprehension and cognitive loads.
Specifically, Li et al.\cite{li2015novel} point out that two charts of the same data but with different layouts can cause perceptional imbalance, even though they are informatively equivalent.
Despite the importance, the design space of chart attributes is too large for designers to choose from, resulting in various types of charts in designers' own style\cite{shi2022supporting}.
Most of these works follow a coarse-to-fine strategy to first identify chart types and then extract visual marks and channels~\cite{savva2011revision, jung2017chartsense, poco2017extracting, yuan_2021_deep}. 

Existing methods often mix visual attributes as global style for similarity estimation~\cite{ma2018scatternet,siegel2016figureseer,jung2017chartsense}, in which not all attributes are of interest to users~\cite{li2022structure}.
To fill the gap, the \textit{Annotation} stage of the proposed WYTIWYR framework allows users to disentangle attributes in a given chart.

\noindent
\textbf{Chart Retrieval}.
The core for effective retrieval is to delineate the similarity between the query input and retrieval candidates in the database.
Based on the object of similarity measurement in the retrieval process, existing chart retrieval strategies can be mainly categorized into two classes: definition- and example-based approaches.
Definition-based approaches (\eg,~\cite{chen2015diagramflyer, siegel2016figureseer, hoque2019searching, zeng_2021_vistory, zhao_2022_chartseer,li2022structure}) characterize the criteria of similarity by explicit chart attributes, making it preferable for well-configured chart formats, including SVG-style and Vega-Lite grammars.
For instance, Hoque and Agrawala\cite{hoque2019searching} developed a search engine for D3 visualizations by using a JSON-like configuration that dictates query visual encoding.
Moreover, some studies (\eg,~\cite{heer2008graphical, stitz2018knowledgepearls}) further use user interaction records and visualization states represented in a provenance graph to match the previous exploration states.
Recent example-based methods regard visualization charts as images and leverage deep learning models~\cite{ma2018scatternet, zhang2021chartnavigator,siegel2016figureseer,jung2017chartsense,ye_2022_visatlas} to automatically extract visual features via an end-to-end manner.
A set of implicit low-level features are leveraged to estimate the global similarity.
However, definition-based approaches only consider partially predetermined visual attributes, while example-based approaches tend to match the query with all indiscriminate attributes.

Our proposed WYTIWYR framework combines the advantages of example-based approaches in terms of their capability of capturing implicit attributes and also definition-based approaches that leverage explicit attributes with high interpretability.
Moreover, the \textit{Retrieval} stage enables a flexible combination of these attributes with respect to user intent.

\noindent
\textbf{Vision-Language Pretraining}.
Large-scale pretrained models have shown promising performance in both natural language processing and vision tasks these years, such as BERT~\cite{devlin2018bert}, RoBerTa~\cite{liu2019roberta}, and GPT series~\cite{brown2020language, radford2018improving,radford2019language}.
Following the paradigm of pretrain and finetune, many downstream tasks~\cite{goyal2017making, xu2015show, lu202012} transfer the knowledge from the pretrained model without training a new model from scratch by utilizing it.
Prompt, a text sequence in the form of natural language, links the pretrained model and downstream tasks as a bridge.
Strobelt \emph{et al.}~\cite{strobelt2022interactive} develops a visual interface to provide a promising prompt evaluation.
Combined with a high-capacity text encoder and visual encoder, contrastive language-image pretraining (CLIP)~\cite{radford2021learning} learns heterogeneous multi-modality representations from 400 million image-text pairs by resorting to semantic supervision from the embedding space of CLIP. Several studies~\cite{patashnik2021styleclip,gal2021stylegan,gu2021zeroshot} extend its applicability in a zero-shot manner, which means the model can predict samples whose class is not observed during training.

In this work, we take advantage of vision-language pretraining models, to encode user intents as a prompt to collaborate with the decisive process of retrieval.
The CLIP-driven model is leveraged to align user intent with corresponding visual attributes in both \textit{Annotation} and \textit{Retrieval} stages.
In this way, we can offer highly customizable attributes for chart annotation and multi-modal representation extraction for chart retrieval.

\section{WYTIWYR Framework}
\label{sec:framework}

To build the WYTIWYR framework, we first formulate chart attributes based on a preliminary study (Sec.~\ref{ssec:attribute_formulation}), then
present an overview of the WYTIWYR framework (Sec.~\ref{ssec:pipe}). 

\begin{figure}[t]
    \centering
    \includegraphics[width=0.8\linewidth]{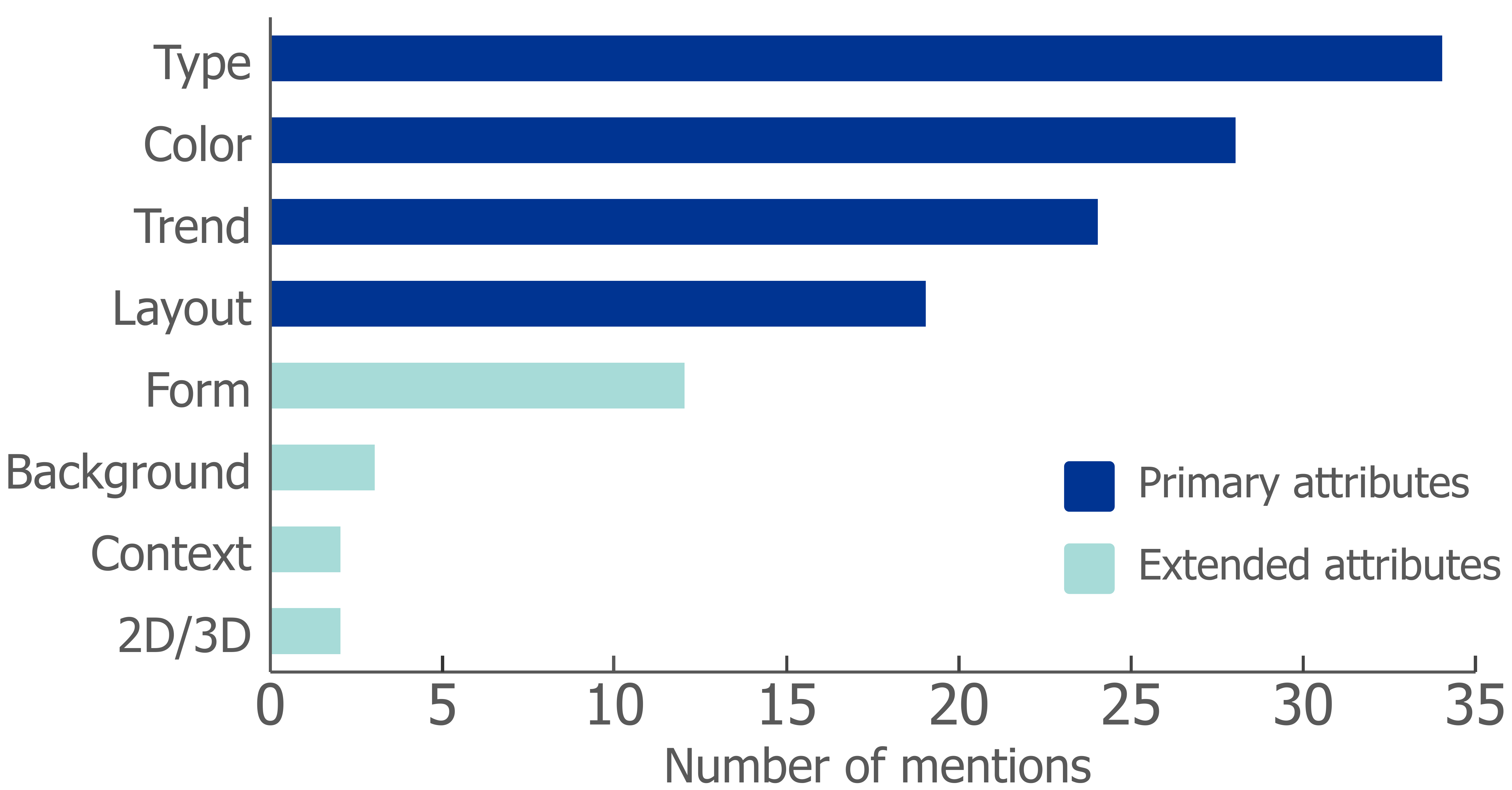}
    \vspace{-3mm}
    \caption{\textbf{Results of our preliminary study.} 
    \emph{Type, color, trend,} and \emph{layout} are  voted by most participants and identified as \emph{primary} attributes, whilst the others are categorized as \emph{extended} attributes.
    }
    \vspace{-4mm}
    \label{fig:Preliminary_attr_type}
\end{figure}

\subsection{Attribute Formulation}
\label{ssec:attribute_formulation}

There are two categories of chart attributes for chart retrieval: 1) \emph{primary attributes} that are generally considered for chart retrieval, and 2) \emph{extended attributes} that meet the specific needs of different users.
To better understand user intents, we conducted a comprehensive study to categorize chart attributes.

\noindent\textbf{Study design.} 
Our preliminary study was conducted online and involved 40 participants whose ages ranged from 20 to 52 (mean = 24.3).
Before the study, we first introduced the background and purpose of the experiment to the participants. 
Then we tested their knowledge on the chart through a set of verification questions. 
Only the participants who could identify typical charts, such as bar and pie charts, were allowed to participate in the study.
Fortunately, all participants were familiar with charts since their popularity on social media.
Next, the participants were shown multiple real-world charts and asked to indicate what attributes were of interest.
\revise{To eliminate any potential subjective bias during the chart selection process for testing, a random sample of five images (three in the Beagle dataset\cite{battle2018beagle}, two in other real-world examples) was taken for 18 types of chart from the collected data.}
The testing charts were carefully chosen according to chart type taxonomy as in the study by Borkin \textit{et al.}\cite{borkin2013makes}.
\revise{The detail of sampled testing chart can be seen at supplementary material.}
Their answers were recorded in text and summarized in different perspectives after the experiments.
The perspectives were cross-validated by two of our co-authors to ensure correctness.

\begin{figure}[t]
    \centering
    \includegraphics[width=0.9\linewidth]{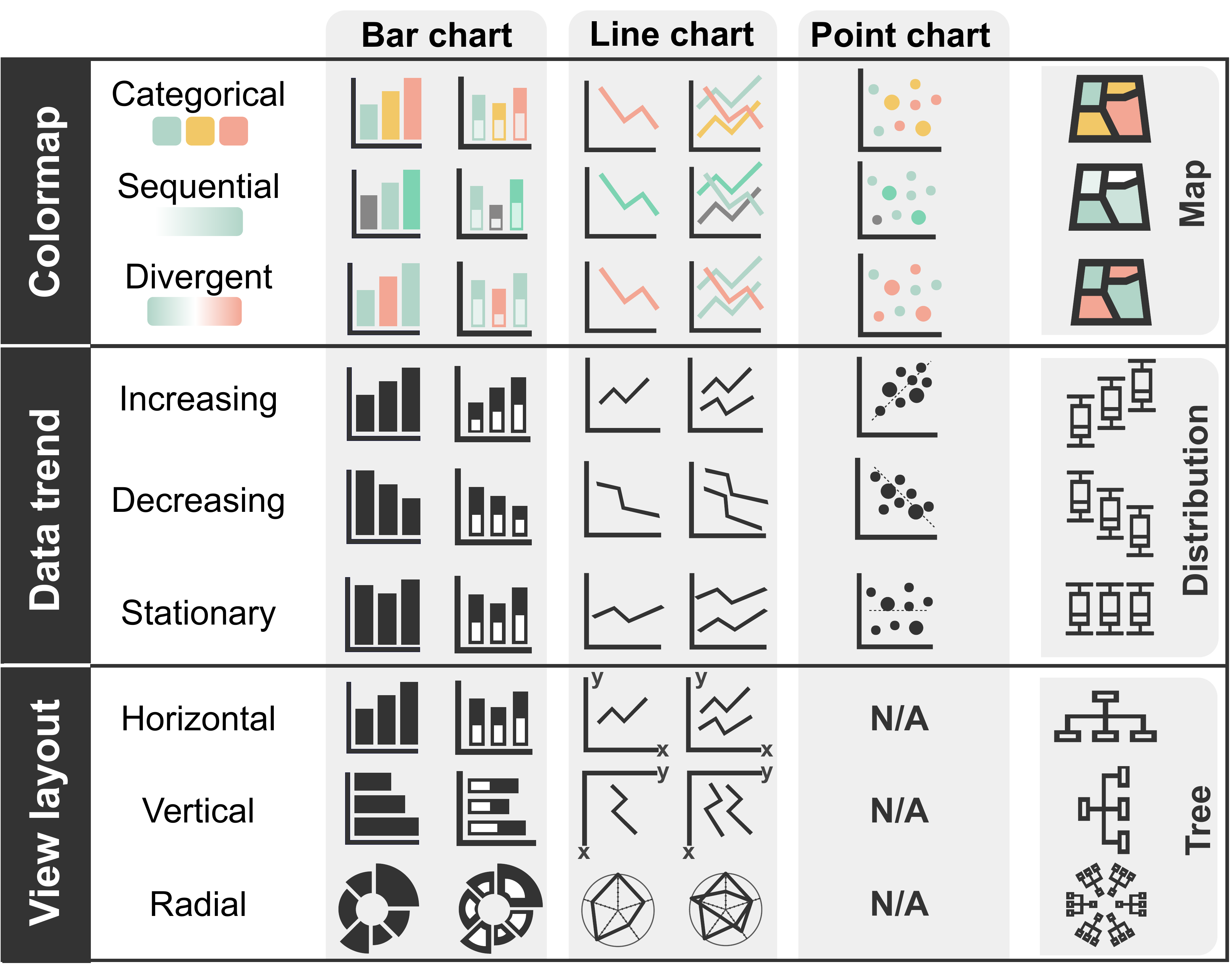}
    \vspace{-2mm}
    \caption{\textbf{Illustration of attribute options of primary attributes for several chart types.} Some options may not be available for certain chart types, such as \emph{Layout} for the point chart. 
    }
    \label{fig:Taxonomy_out}
    \vspace{-6mm}
\end{figure}

\begin{figure*}[t]
\centering
\includegraphics[width=0.9\linewidth]{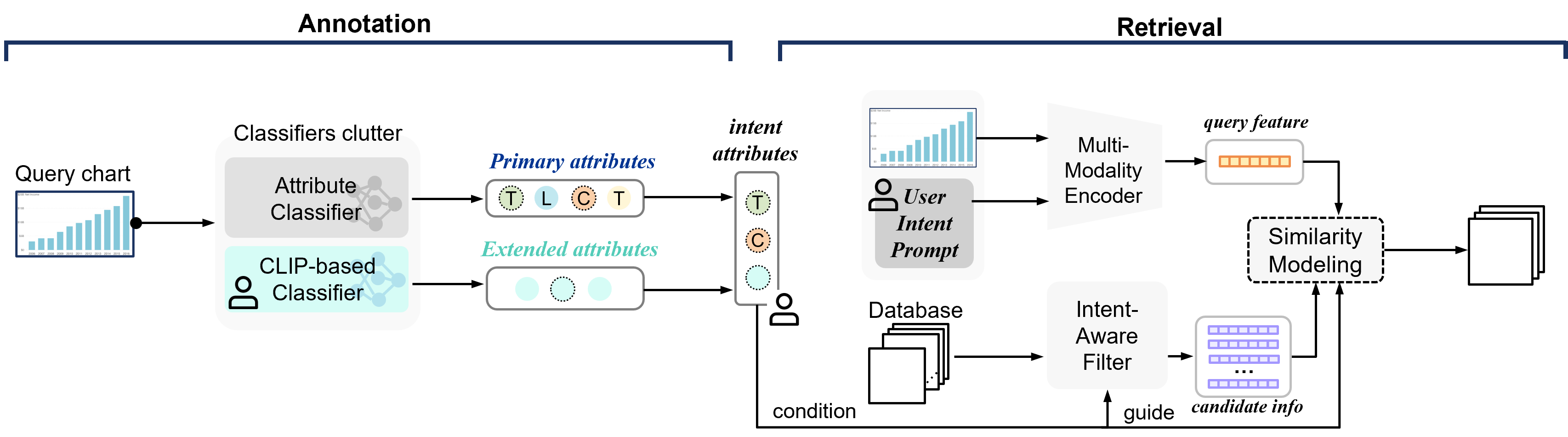} 
\caption{\textbf{Overall pipeline of attributes-aware retrieval.} The pipeline consists of two stages: Annotation for extracting the attributes and Retrieval for modeling the similarity between query charts and charts in the database.}
\vspace{-4mm}
\label{fig: overview}
\end{figure*}

\noindent\textbf{Result.}
As shown in Figure~\ref{fig:Preliminary_attr_type}, we identify a total of eight perspectives of attributes are of interest from the feedback, which are also consistent with previous studies~\cite{borkin2013makes,li2022structure}.
Among them, \emph{Type, Color, Trend}, and \emph{Layout} are the four attributes mostly frequently mentioned by the participants and can be applied to most types of charts.
\revise{Although the \emph{Form} attribute was also frequently mentioned by participants, it is primarily associated with line charts and less frequently mentioned in the other chart types.
Therefore, to ensure generalization and expressivity, we decided not to combine \emph{Form} with the other four attributes.}
We dub \primary{\textbf{\textit{primary attributes}}} as ${\primary{\mathcal{T}_p}} := \left\{Type,\text{ }Color,\text{ }Trend, \text{ }Layout \right\}$.
In addition, the participants also brought up some other chart attributes, including \emph{form, background, context}, and \emph{2D/3D}, which we dubbed as {\extended{\textbf{\textit{extended attributes}}}} {\extended{$\mathcal{T}_e$}} that includes the other attributes.
For these chart attributes, the variation between different users is large, making it difficult to be comprehensively enumerated.

Moreover, the study also revealed a set of design options for each chart attribute. 
Specifically, we choose 10 primary categories and 18 subcategories of chart types from the visualization taxonomy\cite{borkin2013makes}.
Figure~\ref{fig:Taxonomy_out} illustrates the options in the attribute type of \{\emph{color}, \emph{trend}, \emph{layout}\} for several chart types including bar, line, and point charts.
Notice that some chart types may not have all options mentioned above.
For example, \emph{layout} for point charts are not available, and maps only have \emph{color} options. 

\subsection{Overall Pipeline}
\label{ssec:pipe}

The workflow of our proposed WYTIWYR framework is depicted in Figure~\ref{fig: overview}. 
For the \textit{Annotation} stage, we employ several classifiers based on robust neural network architecture to disentangle $\{\primary{\mathcal{T}_p}, \extended{\mathcal{T}_e}\color{black}{\}\in\mathcal{T}}$ embedded in the query chart $\mathcal{Q}$.
After that, users can select the disentangled attributes $\mathcal{I}_A$ as their intents based on their needs. Also, $\mathcal{I}_A$ would condition the components of the filter in the retrieval stage.
For the \textit{Retrieval} stage, we take a dual-path similarity modeling between the \textit{query feature} and the \textit{candidate info} with the guidance of intent attributes to yield the return charts. For the path of the \textit{query feature}, bitmap query chart $\mathcal{Q}$ with user intent text prompt $\mathcal{I}_P$ is fed into a multi-modal encoder to obtain the joint representation. For the \textit{candidate info} path, the feature is produced by charts in the database filtered by the intent-aware filter built by $\mathcal{I}_A$. 
Throughout the overall pipeline, user intent can be injected and guide the chart retrieval process by the following three ways:

\begin{itemize}
    \item \textbf{Classifier Customization.} 
    Users can tailor the usage of classifiers depending on their needs in order to determine the presence of an attribute in the query chart.
    \item \textbf{Disentangled Attributes Selection.} The attributes adopted in the \textit{Retrieval} stage can be selected and combined by the users.
    \item \textbf{User Prompt Tuning.} Users can add specific implicit intent as the text prompt to guide the retrieval process, which is independent of the query chart.
\end{itemize}
\section{WYTIWYR Prototype System}
\label{sec:prototype}

\subsection{Stage1: Annotation}
\label{ssec: annotation}

As shown in Figure~\ref{fig: annotation}, the task of the \textit{Annotation} stage is to disentangle and generate the visual attributes $\mathcal{T}$ adhered to the given query chart $\mathcal{Q}$.
For the primary attributes $\primary{\mathcal{T}_p}$, we adopt four classifiers for separate extraction in a supervised learning manner. 
For the extended attributes $\extended{\mathcal{T}_e}$, as they are absent from our training, we conduct the extraction process in the fashion of zero-shot learning. 


\subsubsection{Annotation for Primary Attributes}
\label{sssec: Annotation for Requisite Attributes}
\noindent\textbf{Redundant Information Removal.}
Several practical scenario-based raw data are shown in Figure~\ref{fig: bg_rm} (a), including fancy decoration, text information, background, and legends. 
This work focuses on the intrinsic attributes of $\mathcal{Q}$ instead of this redundant information.
Hence we remove the redundant information using ISNet~\cite{qin2022highly},  a segmentation network with pretrained weights.

Nevertheless, there exists context beyond the segmentation capability of ISNet, \revise{such as the emojis in the left-most example in Figure \ref{fig: bg_rm} (a) \& (b)}. 
\revise{The redundant emojis in red and yellow colors} would degrade the performance of colormap classification.
Instead of being categorized as 
\textit{categorical colormap}, the chart would be mistakenly recognized to \textit{diverging colormap}.
To mitigate such issue, we ignore the colors that share less than 10\% pixels of non-transparent regions of the image after segmentation.

\begin{figure}[t]
\centering
\includegraphics[width=0.925\columnwidth]{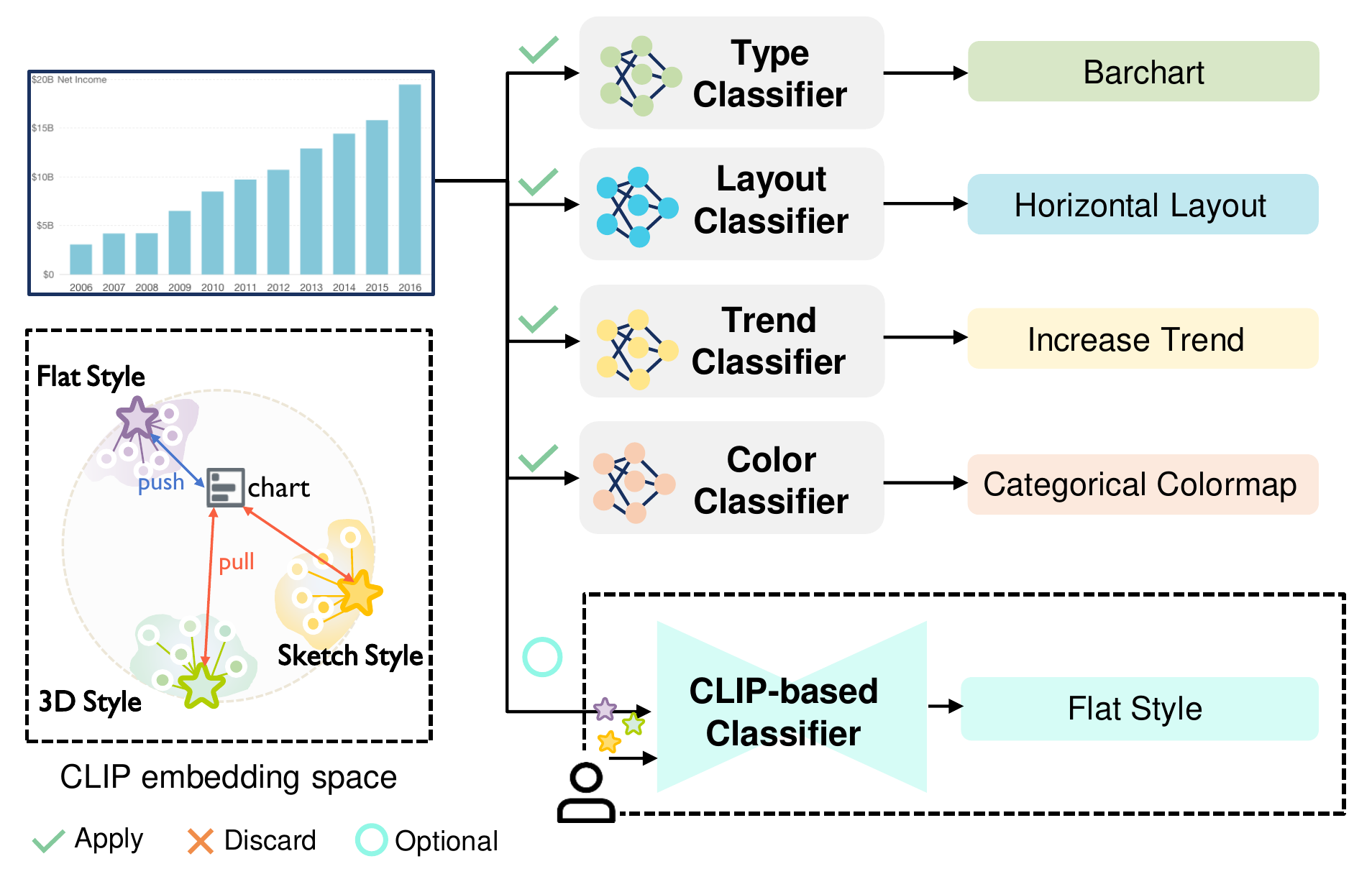}
\vspace{-3mm}
\caption{\textbf{The \textit{Annotation} stage of our pipeline.} To disentangle various attributes in the query chart, four classifiers are built for primary attributes. In addition, one CLIP-based classifier is employed to optionally annotate the extended attribute via user customization.}
\vspace{-5mm}
\label{fig: annotation}
\end{figure}

\begin{figure}[t]
\centering
\includegraphics[width=0.95\columnwidth]{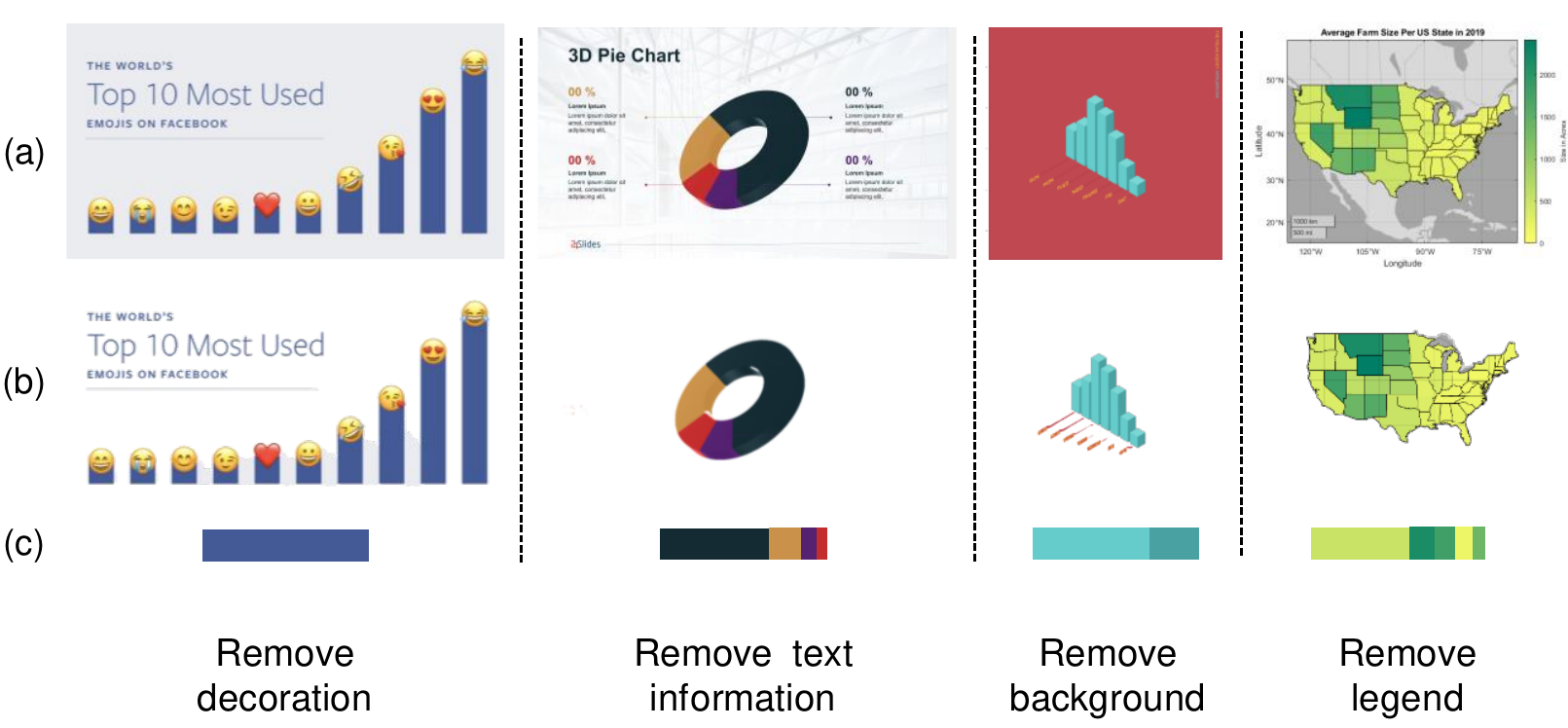} 
\vspace{-3mm}
\caption{\textbf{Redundant information removal and main color extraction.} (a) Raw images. (b) The chart after removing redundant information. (c) The extracted color after filtering redundant colors.}
\vspace{-4mm}
\label{fig: bg_rm}
\end{figure}

\noindent\textbf{Unified Classifiers.}
For annotation of the primary attributes, we employ ResNet50\cite{he2016deep} architecture as the backbone for \revise{all} four attribute classifiers. Thanks to the residual connection among the convolution layers, the network is both lightweight and performance-guaranteed. 
During the training process, the network will learn the complex features under the supervision of given attributes.
In the reference process, the trained classifier would identify the corresponding attribute in the query chart, as in Figure~\ref{fig: annotation}.

The only difference in the four classifiers is the number of output channels in the last fully connection layer. For instance, the number of $Type$ classifiers is set to $18$ since there are $18$ types of charts under consideration. Similarly, based on the specific task, we set this number to $3$ for the $Layout$, $Trend$, and $Color$ classifiers with the respect to their attribute options; see Figure~\ref{fig:Taxonomy_out} for detail.
Note that $\mathcal{Q}$ may not have the coverage of all attributes included in the classifiers. For instance, $\primary{\mathcal{T}_p}$ of a heatmap chart is $\left\{Type,\text{ }Color\right\}$ while $\left\{Trend,\text{ }Layout\right\}$ attributes are absent. 

\noindent\textbf{Loss Function.}
There are two challenges to overcome when formulating the loss function: 1) imbalanced samples in each class of the dataset (see Table~\ref{tab1: dataset}), and 2) the existence of noisy samples that are hard to classify.
In order to improve the accuracy for $\primary{\mathcal{T}_p}$ annotation, we introduce Focal Loss~\cite{lin2017focal}, which modifies the standard cross entropy loss to overcome the above-mentioned challenges.
The Focal Loss is defined as:
\begin{equation}
\label{eq2}
	\mathcal{L_{\mathrm{FL}}}(p_{t}) = -\alpha_{t}(1-p_{t})^{\gamma}\operatorname{log}(p_{t}),
\end{equation}
where $p_{t}\in[0, 1]$ represents the estimated probability of class $t$, 
$\alpha_{t}$ represents the scaling factor, and $\gamma$ represents the modulating factor. Among them, $\alpha_{t}$ is set by inverse class frequency, thus learning parameters tend to contribute to classes with fewer samples, and
$\gamma$ assists in up-weight the loss assigned to poor-classified examples, avoiding the possibility that the training process is dominated by the amount of well-classified samples.

\subsubsection{Annotation for Extended Attributes}
\label{sssec: Annotation for Intent Attributes}
Since the primary attributes only consider the general needs in chart retrieval, we offer an optional classifier manipulated by the users to distinguish {\extended{$\mathcal{T}_e$}} in this query chart, as shown in Figure~\ref{fig: annotation}.
The input of {\extended{$\mathcal{T}_e$}} classifier requires the user to provide several labels of the attribute apart from $\mathcal{Q}$ (\eg, style). We denote these labels as $\{{t}_1, {t}_2, \ldots, {t}_m\}\in {T}$. As these labels are out of our dataset, a task of zero-shot classification is naturally formed.
In the following, we will briefly introduce the mechanism of the CLIP model at a general level with a toy example as in Figure \ref{fig: annotation}, 
where the user sets the text labels as [\textit{``3D style''}, \textit{``Flat style''}, \textit{``Sketch style''}].

The CLIP model aligns $T$ embeddings with $\mathcal{Q}$ embeddings in a multi-modality embedding space in a contrastive manner. 
Specifically, in the embedding space of the example in Figure~\ref{fig: annotation}, the distance of $\mathcal{Q}$ and \textit{``3D style''} would be less than the other two irrelevant labels.
\revise{As such, the query results tend to be \textit{``3D style''} bar charts.
The CLIP model balances well between the performance and computational cost in the prototype system.
Nevertheless, the model in our framework can be feasibly replaced with other advanced language-image models, \eg,\cite{li2022blip, li2021align, zhang2022tip}.}


\subsection{Stage2: Retrieval}
\label{ssec: retrieval}

\begin{figure}[t]
\centering
\includegraphics[width=\columnwidth]{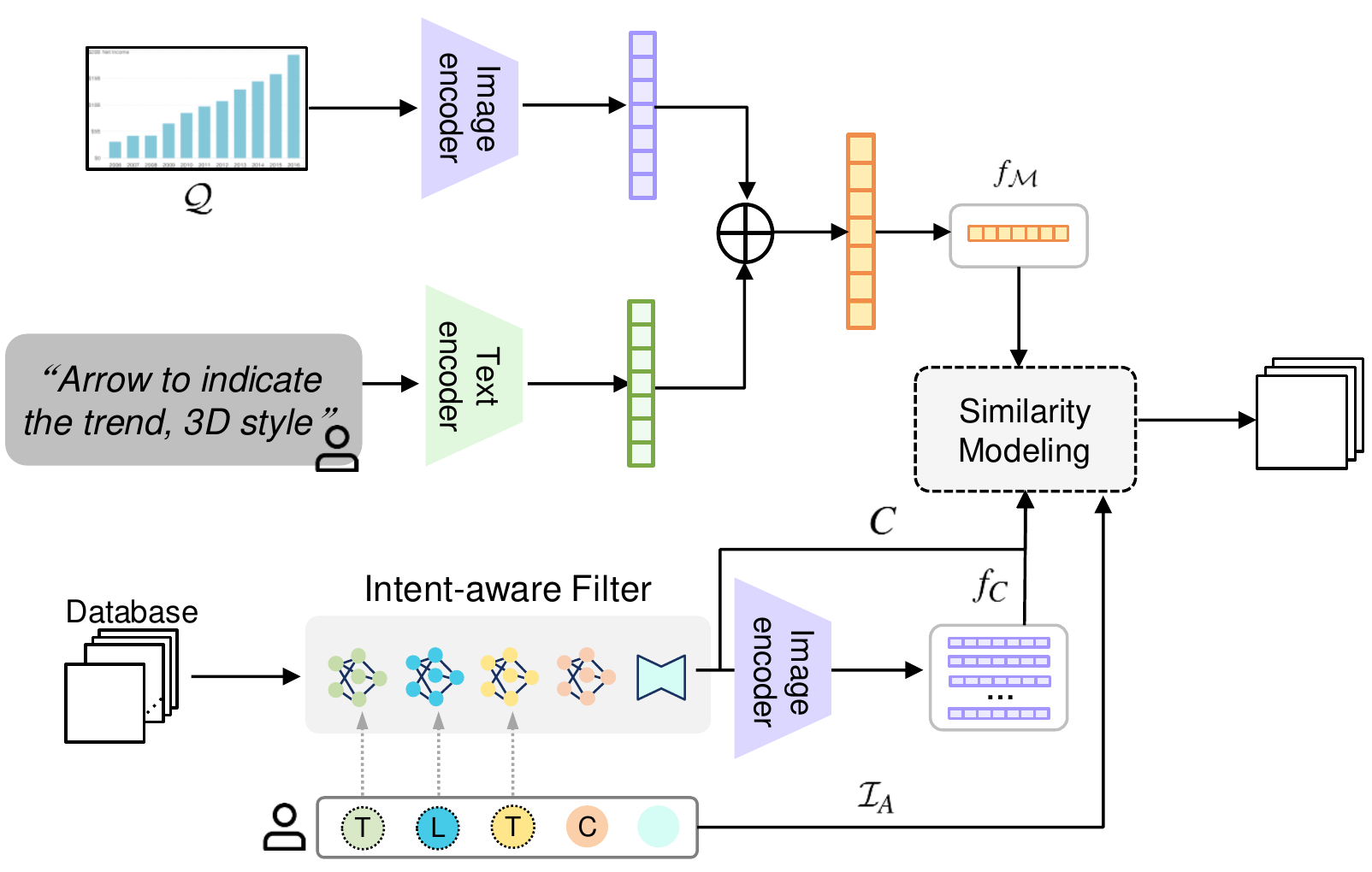} 
\caption{\textbf{The \textit{Retrieval} stage consists of two branches.} The \revise{intent-aware filter branch (bottom)} filters out charts based on intent attributes selected by the user, generating multiple chart candidates. The \revise{multi-modal encoder branch (top)} produces the query feature based on the multi-modal inputs. Similarity modeling is conducted between results of these two branches.}
\vspace{-4mm}
\label{fig:retrieval}
\end{figure}

As the \textit{Annotation} stage completes the disentanglement of the attributes $\mathcal{T}$ in the query chart, users in the \textit{Retrieval} stage can select the intent attributes $\mathcal{I}_A$ based on their \revise{intents}.
Depicted in Figure~\ref{fig:retrieval}, there are two branches with different modules for our \textit{Retrieval} stage, namely multi-modal encoder and intent-aware filter.
The intent-aware filter rules out charts whose attributes are beyond the range of $\mathcal{I}_A$.
The multi-modal encoder integrates $\mathcal{Q}$ and $\mathcal{I}_P$ as a joint representation.
Then, similarity modeling is conducted between \revise{results of} these two branches, yielding the final retrieval results based on the modeling score.

\subsubsection{Multi-Modal Encoder}
This branch would form a CLIP-generated \textit{query feature} with inputs $\mathcal{Q}\text{, }\mathcal{I}_P$. As introduced in Sec.~\ref{sssec: Annotation for Intent Attributes}, the CLIP embedding space has heterogeneous aligned text and visual features from the multi-modal input. 
Hence, $\mathcal{Q}$ and $\mathcal{I}_P$ are encoded into features by their respective CLIP encoders, and fused as a joint multi-modal feature $f_{\mathcal{M}}$ that can be denoted as:
\begin{equation}
\label{eq3}
	f_{\mathcal{M}} = \frac{F_{\theta}(\mathcal{Q}) + G_{\phi}(\mathcal{I}_P)}{2},
\end{equation}
where $F_{\theta}$ and $G_{\phi}$ are denoted as the image encoder and text encoder of the CLIP model, respectively.

\subsubsection{Intent-Aware Filter}
The branch of Intent-Aware Filter rules out irrelevant charts in the database and forms the \textit{candidate feature}.
Previous works (\eg,~\cite{hoque2019searching, ma2018scatternet}) on chart filtering primarily rely on fixed settings of the model, which are coarse and neglect the user's \revise{intents}.
For this shortcoming, we propose an intent-aware filter constructed by $\mathcal{I}_A$, a set of disentangled attributes selected by the user.
For example, with $\mathcal{Q}$ illustrated in Figure~\ref{fig:retrieval}, a user may select the [\textit{Type}, \textit{Layout}, \textit{Trend}] as the target attributes for the retrieval, then the filter retains $n$ charts containing these target attributes in the chart database.
Then, the retained charts would serve as candidates $C:=\{c_{1}, c_{2},\cdots, c_{n}\}$ that are encoded as \textit{candidate feature} $f_{C}:=\{f_{c_{1}}, f_{c_{2}},\cdots, f_{c_{n}}\}$ by image encoder in the CLIP model.

\subsubsection{Similarity Modeling}
As the core of retrieval, similarity modeling is comprehensively performed with triplets $\{\mathcal{Q}$\text{, }$\mathcal{I}_A$\text{, }$f_{\mathcal{M}}\}$. Among them, $\mathcal{Q}$ provides the global perception composed of implicit features.  
The explicit intent attributes $\mathcal{I}_A$ and implicit query feature $f_{\mathcal{M}}$ that include user intent prompt and query chart, work together to recall user-intended examples throughout the retrieval process.
The overall similarity score $\mathcal{S}$ can be denoted as:
\begin{equation}
\label{eq4}
	\mathcal{S} = S_{\mathcal{Q}} \cdot \text{e}^{\nu  S_{\mathcal{I}_A} +\mu  S_{\mathcal{M}}},
\end{equation}
where $S_{\mathcal{Q}}$ demotes the global perception at pixel level, and $S_{\mathcal{I}_A}$ and $S_{\mathcal{M}}$ denotes the similarity score of user-selected attributes and the multi-modality feature, respectively. The scaling factors $\nu$ and $\mu$ are empirically set as $1$ and $5$, respectively. All $S_{\mathcal{Q}}$, $S_{\mathcal{I}_A}$ and $S_{\mathcal{M}}$ are normalized with range of $[0, 1]$. In the following, we will introduce these similarity scores in detail.

\noindent \textbf{Global Perception Score.} Although salient attributes with respect to user intent are disentangled and selected, a multitude of implicit features are intertwined within the chart image, which forms the concept of global perception.
We adopt the classic cosine similarity and compute it at the pixel level between $\mathcal{Q}$ and the feature of every candidate $f_{C_{i}}, i=1, 2, \ldots, n$ as follows: 
\begin{equation}
\label{eq4}
	S_{\mathcal{Q}} = \frac{{ F_\theta({\mathcal{Q}}}) \cdot {f_{C_{i}}}}  {\|{F_\theta({\mathcal{Q}}})\| \|{f_{C_{i}}}\|}.
\end{equation}

\noindent \textbf{Intent Attributes Score.}
Extended attribute $\extended{\mathcal{T}_e}$ contains a user's expected attributes of retrieval results charts apart from four primary attributes. 
Among the attributes, \textit{Type} and \textit{Layout} define core properties of a chart, which is easy to distinguish from $\mathcal{Q}$. Therefore, we neglect them in the similarity crafting. The other attributes serve as the miscellaneous variant, and the change of them can be minuscule. This motivates us to build another score $S_{\mathcal{I}_A}$ to further enhance the retrieval process.
Therefore, we mainly consider the following three attributes: \textit{Trend} $\mathcal{N}$, \textit{Color} $\mathcal{C}$ and \textit{extended attribute} $\extended{\mathcal{T}_e}$.
$S_{\mathcal{I}_A}$ is formulated as:
\begin{equation}
\label{eq5}
	S_{\mathcal{I}_A} = S_{\mathcal{N}} + S_{\mathcal{C}} + S_{\mathcal{T}_e}.
\end{equation}

\noindent For $S_{\mathcal{N}}$, we build an extractor to extract the trend feature from the query chart $\mathcal{Q}$ and every candidate chart $C_{i}, i= 1, 2, \ldots, n$.
The extractor shares the parameters with the trend classifier in the \textit{Annotation} stage except the last fully connected layer.
Then, $S_{\mathcal{N}}$ can be estimated by cosine similarity between the extracted features.

For $S_{\mathcal{C}}$, we follow the scheme in Figure~\ref{fig: bg_rm} to transform $\mathcal{Q}$ into a proportional color palette after the steps of background removal and color extraction. 
Then the proportional color palette is transformed into a 128-bin color histogram in all RGB channels and separately stored in three vectors. We then concatenate these vectors to form the ultimate color vector $V$. Denote $V_{\mathcal{Q}}$, $V_{C_{i}}, i= 1, 2, \ldots, n$ as the color vectors of $\mathcal{Q}$, $V_{C_{i}}, i= 1, 2, \ldots, n$, respectively, we then estimate the cosine similarity between them.

For $S_{\mathcal{T}_e}$, since it is hard to quantify the similarity of intents, we leverage the power of the CLIP model again to give an accurate response. We feed the CLIP model with candidate chart $C_{i}, i= 1, 2, \ldots, n$ and text labels in user intent $t_i, i= 1, 2, \ldots, m$. For each chart, we would obtain $m$ outputs $\{y_1, y_2, \ldots, y_m\}$. Prior to the computation, the user would select one text label indexed $s$ as the attribute that best represented their intent. Then, we denote $S_{\mathcal{T}_e}$ for a candidate chart as:
$ S_{\mathcal{T}_e} = \text{e}^{y_{s}}/{\sum_{m}\text{e}^{y_{m}}}.$

\noindent \textbf{Feature Matching Score.} Similarly, we match the closeness between the multi-modal feature and the candidate feature by the following equation:  
$S_{\mathcal{M}} = \frac{f_{\mathcal{M}} \cdot {f_{C_{i}}}} {{\|{f_{\mathcal{M}}\| \|{f_{C_{i}}}\|}}}.$


\subsection{System Interface}
We design an interface for the prototype system that allows users to retrieve a chart according to their intents.
\begin{figure}[t]
    \centering
    \includegraphics[width=0.98\linewidth]{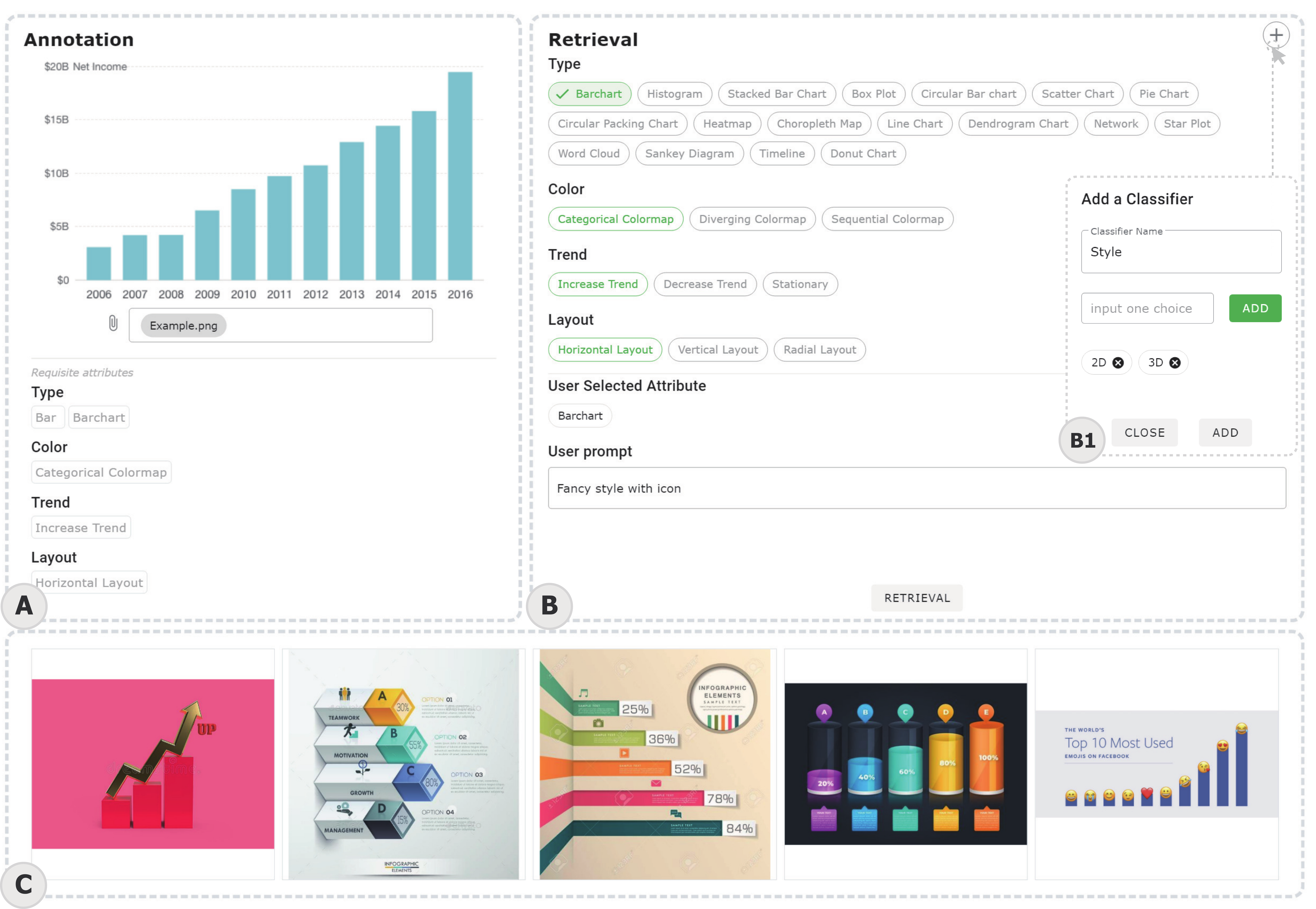}
    \caption{\textbf{Interface of our prototype system}. A) the \emph{Annotation} view decouples the properties of the query chart; B) the \emph{Retrieval} view allows users to select the primary and extended attributes, and C) the \emph{Result} view shows the top 5 retrieved charts.}
    \label{fig:Interface}
    \vspace{-4mm}
\end{figure}

\noindent\textbf{Annotation View. }
In this view, users can upload a query chart and get disentangled primary attributes, including \textit{Type, Color, Trend}, and \textit{Layout}.
As shown in Figure~\ref{fig:Interface} (A), a bar chart is uploaded and the automatically annotated attributes, ``\emph{Barchart, Categorical Colormap, Increasing Trend, Horizontal Layout}'' are presented.

\noindent\textbf{Retrieval View.}
In this view (Figure~\ref{fig:Interface} (B)), users can view and choose both primary and extended attributes. 
The first tag in green corresponds to the annotated attributes from the query chart, while other options are also shown, allowing users to select the desired attributes. 
If users feel that the default primary attributes are insufficient or have other attributes of interest, they can add a new classifier by clicking on the chart in the upper right corner, which reflects the philosophy of the proposed intent-aware design.
In the input box below, users can enter their intentions as intent prompts for more customized queries.
As shown in Figure~\ref{fig:Interface} (B1), the user selects ``Bar chart'' for the type attribute and enters \prompt{``Fancy style with icon''} in the Retrieval view.

\noindent\textbf{Result View.}
This view (Figure~\ref{fig:Interface} (C)) shows top five query results, arranged in decreasing order according to their similarity scores.

\section{Evaluation}
\begin{figure*}
    \centering
    \includegraphics[width=0.95\linewidth]{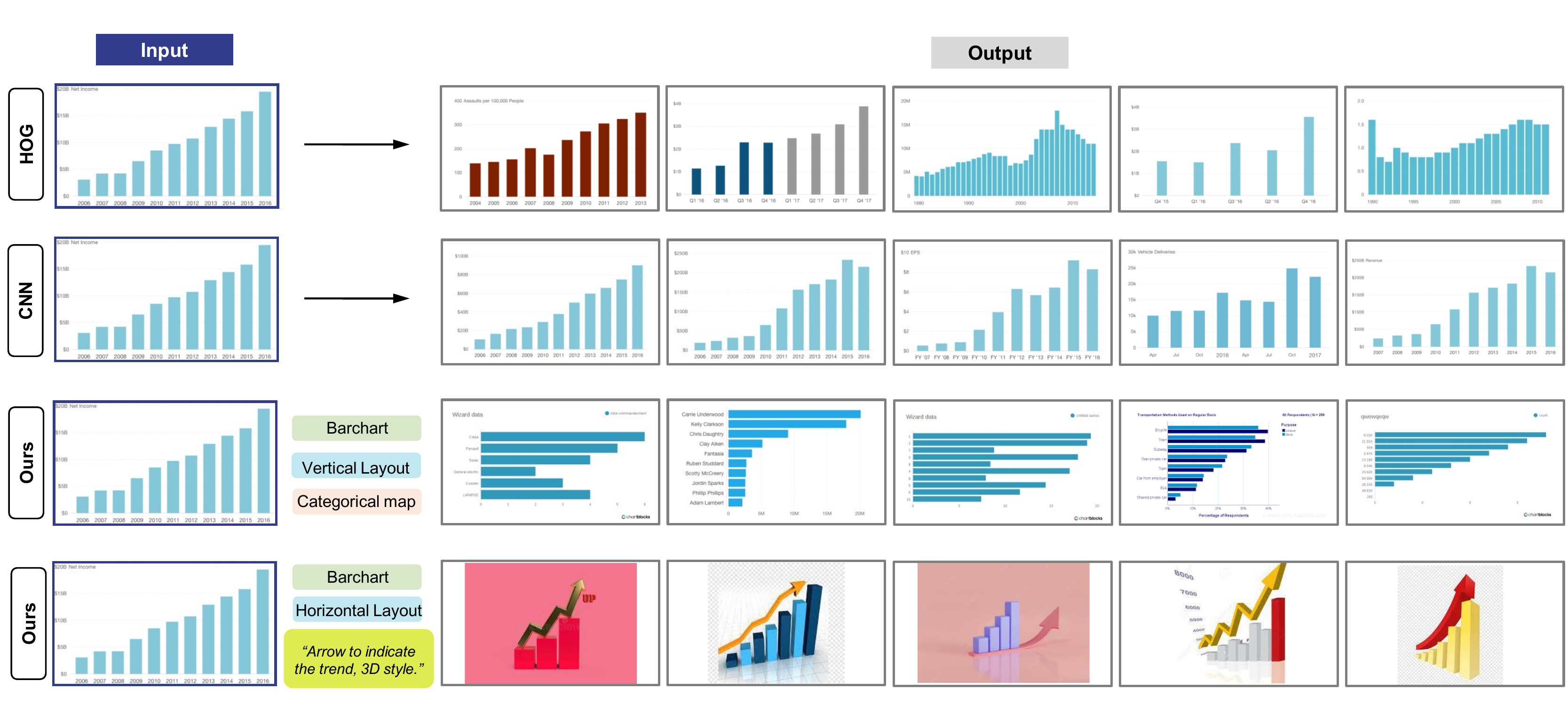}
\vspace{-3mm}
    \caption{\textbf{Visual comparison of chart retrieval by different methods.} First row: Retrieval results by the conventional HOG method. Second row: Retrieval results by the deep-learning-based CNN method. Third to fourth rows: Retrieval results of our WYTIWYR framework.}
    \label{fig:experiment}
\vspace{-3mm}
\end{figure*}

\begin{table}[h]
\centering
\fontsize{8}{11}\selectfont
\caption{Our integrated dataset \revise{consists of 18 types of charts} from Beagle\cite{battle2018beagle} and manual collection.}
\label{tab1: dataset}
\begin{tabular}{|l|c||l|c|}
\hline 
\rowcolor{Gray}
\multicolumn{1}{|c|}{Chart Type} & \multicolumn{1}{c||}{\#Count} & \multicolumn{1}{c|}{Chart Type} & \multicolumn{1}{c|}{\#Count} \\ \hline \hline
Bar chart    &7269   & Heatmap    & 352\\
Stacked Bar Chart   &1159   & Line Graph  &11605\\
Circular Bar Chart   &608   & Star Plot   &491\\
Donut Chart    &2459   & Choropleth Map   &640\\
Pie Chart   &2587   & Scatter Plot  &3000\\
Sankey Diagram   &266   & Word Cloud     &406\\
Timeline   &324   & Dendrogram    &298\\
Box Plot    &571   & Network    &395\\
Histogram    &695   & Circular packing chart  &139\\

\hline 
\end{tabular}
\vspace{-2mm}
\end{table}

\subsection{Quantitative Experiment}
\label{ssec:Quantitative}

\noindent
\textbf{Dataset.}
Previous works concentrate on retrieval of limited chart types, \textit{e.g.}, single~\cite{ma2018scatternet} or a few type~\cite{chen2015diagramflyer, hu2019vizml} chart retrieval.
Moreover, some of the literature coarsely considers the categorization of charts, by broadly grouping several chart categories into one~\cite{li2022structure}. 
Many datasets are synthesized with simple composition and monotonous variation, making it difficult to adapt to real-world scenarios that are much more complex.
In this work, we utilize the Beagle dataset~\cite{battle2018beagle} \emph{with charts} in bitmap format, which offers visualization collections designed by real-world users through multiple tools, including D3, Plotly, Chartblocks, Fusion Charts, and Graphiq. 
To make our framework more robust, we further add more difficult instances by manually collecting 4k images from Pinterest, which supplements a large number of stylized and irregular charts.
We filter out charts whose type is beyond our scope (\eg, scientific visualizations) and manually label them with $\mathcal{T}_p$.
Finally, our dataset consists of 33,260 images in total.
The detailed distribution is shown in Table~\ref{tab1: dataset}.

\begin{table}[t]
\centering
\fontsize{8}{11}\selectfont
\caption{Accuracy of \textit{Annotation} results.}
\label{tab2: annotation}
\begin{tabular}{|c||c|c|c|c|}
\hline 
\rowcolor{Gray}
\multicolumn{1}{|c||}{Method} & \multicolumn{1}{c|}{\textit{Type}} & \multicolumn{1}{c|}{\textit{Trend}} & \multicolumn{1}{c|}{\textit{Layout}} &
\multicolumn{1}{c|}{\textit{Color}}\\ \hline \hline
ResNet50+MSE Loss  &0.9518  & 0.8290  &0.9537   & 0.7963\\
ResNet50+Focal Loss   & \textbf{0.9601}   & \textbf{0.8424}   &\textbf{0.9653}   & \textbf{0.8019}\\
\hline 
\end{tabular}
\vspace{-2mm}
\end{table}

\noindent \textbf{Annotation Accuracy.} We evaluate the performance of $\primary{\mathcal{T}_p}$ annotation using ResNet50 architectures with different losses. As Table~\ref{tab2: annotation} lists, the best result is achieved with the designated Focal Loss.

\begin{table}[h]
\centering
\fontsize{8}{11}\selectfont
\caption{F1-scores of \textit{Retrieval} results.}
\label{tab3: retrieval}
\begin{tabular}{|c|c|c|c|c|c|}
\hline 

\cellcolor[gray]{0.9}  &  \cellcolor[gray]{0.9}  & \multicolumn{4}{c|}{\cellcolor[gray]{0.9}F1-Score}\\
\cmidrule{3-6} 
\multirow{-2}{*}{\cellcolor[gray]{0.9}Top-K} & \multirow{-2}{*}{\cellcolor[gray]{0.9}Method} & \multicolumn{1}{c}{\cellcolor[gray]{0.9}\textit{Type}} &\multicolumn{1}{c}{\textit{\cellcolor[gray]{0.9}Trend}} & \multicolumn{1}{c}{\cellcolor[gray]{0.9}\textit{Layout}} & \multicolumn{1}{c|}{\cellcolor[gray]{0.9}\textit{Color}}

 \\ 
 \hline \hline
\multirow{3}{*}{3} & HOG   & 0.6199  & 0.6140   & 0.5449   & 0.6800     \\
                & CNN   & 0.9154  & 0.8043   & 0.7095   & 0.7717       \\
                & Ours    & \textbf{0.9549}  & \textbf{0.8360}   & \textbf{0.9280}   & \textbf{0.8260}     \\ \hline
\multirow{3}{*}{5} & HOG    & 0.5067  & 0.4853   & 0.3857   & 0.6045     \\
                & CNN    & 0.8872  & 0.7550   & 0.6324   & 0.7076      \\
                & Ours   & \textbf{0.9364}  & \textbf{0.7944}   & \textbf{0.9118}   & \textbf{0.7730}     \\ \hline
\multirow{3}{*}{10} & HOG   & 0.4124  & 0.4261   & 0.2741   & 0.5388      \\
                & CNN     & 0.8607  & 0.7085   & 0.5527   & 0.6520     \\
                & Ours   & \textbf{0.9091}  & \textbf{0.7546}   & \textbf{0.8812}   & \textbf{0.7223}     \\
\hline 
\end{tabular}
\vspace{-4mm}
\end{table}

\noindent \textbf{Retrieval Results.}
To evaluate the retrieval performance comprehensively, we examine both precisions and recall via the F1-score.
Denoting the classification  result with true positive, true negative, false positive, and false negative conditions as $TP, TN, FP, FN$, F1-score can be formulated as:
\begin{equation}
\label{eq9}
	F1\text{-}score = \frac{2TP}{2TP+FP+FN}.
\end{equation}

The retrieval performance  with $\mathcal{T}_p$ is individually computed using F1-score in the top-$K$ fashion, where $K \in \{3, 5, 10\}$. 
We devise two settings of with and without user intent to fully evaluate our approach.
For the setting of the absence of user intent, similarity estimation depends on the global perception $S_{Global}$.

To demonstrate the effectiveness of our approach, we take two other approaches as the counterpart: a conventional method (Histogram of Oriented Gradients~\cite{dalal2005histograms}, denoted as HOG from hereon), and a deep-learning-based method (ResNet50, denoted as CNN from hereon).
The quantitative results are displayed in Table~\ref{tab3: retrieval}.
The results reveal that our method significantly outperforms the others.
Furthermore, as $K$ increases, our method keeps the performance superiority while other methods degrade dramatically.
For the evaluation of user intent retrieval, as there are no ground-truth labels to compute quantitative results, we visually compare results generated by different methods for several typical retrieval results.
As shown in Figure~\ref{fig:experiment}, CNN outputs visually more similar results than HOG.
Our method also produces similar results without user intent.
Nevertheless,
our method allows users to add their intent by selecting the disentangled primary attributes ${\color{green}{\mathcal{T}_p}}$ or adding extra intent, which contributes to retrieving results of interest as shown in the last two rows.

\subsection{Case Study}
\label{ssec:Case}
\revise{During the process of visualization design, many inspirations may arise, but bringing those inspirations to fruition can be time-consuming. Retrieval is a fast way to validate those inspirations.}
Inspired by~\cite{bako2022understanding}, two usage scenarios of chart retrieval are illustrated as a proof-of-concept of the WYTIWYR framework: 1) extending the design space by explicit goals and 2) fuzzy retrieval based on implicit user intent. 
For each scenario, we compare the results generated by our intent-aware chart retrieval technique with those generated by intent-free chart retrieval technique.

\subsubsection{Design Space Extension by Explicit Goals}
\begin{figure}[t]
    \centering
    \includegraphics[width=0.97\linewidth]{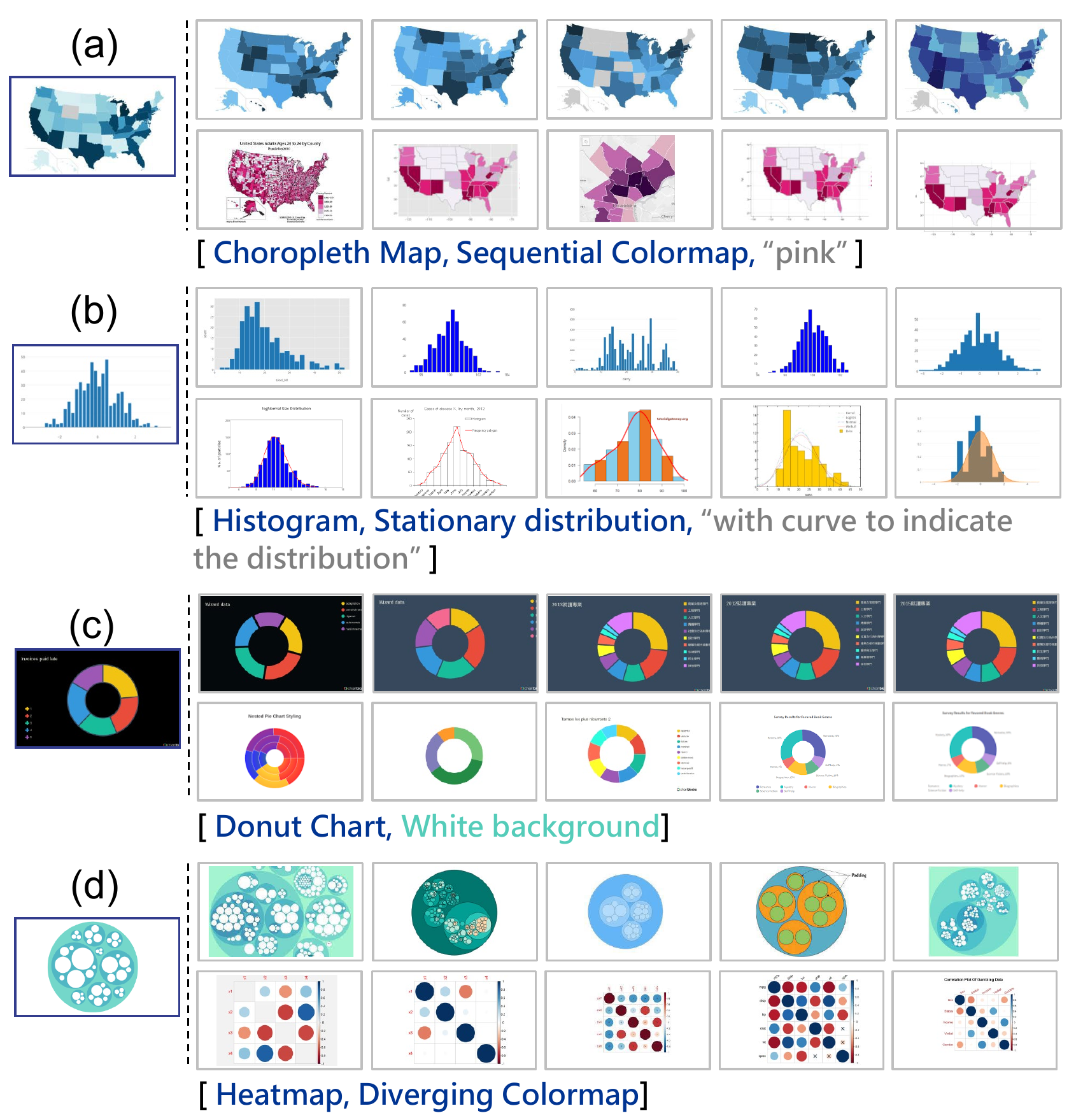}
    \vspace{-3mm}
    \caption{ \textbf{Four cases for extending the design space by explicit attributes.} In each case, the first line presents results without user intent, while the second line presents results with user intent.}
    \label{fig:case1}
    \vspace{-4mm}
\end{figure}

With the prototype system, users can customize retrieval inputs based on disentangled attributes of the query chart and the user intent prompt.
Moreover, beyond the four primary attribute classifiers, users can further add customized classifiers to extract extended attributes.
In this scenario, we divide the disentangled attribute operation into three categories: origin attribute change, new attribute addition, and existing attribute deletion.
In the end, we show an interesting case of attribute transfer that combines these operations.
Figure~\ref{fig:case1} shows four cases of our study.
In general, retrieval without user intent provides similar results as the input but gives minimum design insights.
Instead, with the guidance of user intent, our framework retrieves more diverse results that are well-accepted.
The following lists more details for each case.

\noindent \textbf{Original Attribute Change.}
The disentangled attributes are independent of each other, thus, users can replace one of the attributes while keeping others unchangeable.
In Figure~\ref{fig:case1} (a), the color of the choropleth map is changed to user prompt \prompt{``pink''}. 

\noindent \textbf{New Attribute Addition.} Users can add new attributes, together with the disentangled attributes, to better describe their specific needs.
In Figure~\ref{fig:case1} (b), users add \prompt{``with a curve to indicate the distribution''}, and the results are changed to combinations of histograms and line chart.

\noindent \textbf{Existing Attribute Deletion.} As the attributes can be disentangled as users' needs, they can remove some attributes of no interest. In Figure~\ref{fig:case1} (c), the user dislikes dark background, then can discard the attribute by adding a new classifier with labels of [\textit{``dark background''}, \textit{``white background''}].

\noindent \textbf{Attribute Transfer.} As in Figure~\ref{fig:case1} (d), given a circular packing chart as a query chart, the user gets the attribute annotation as $\{$\emph{Type: ``Circular\; Packing\; Chart'',\; Color: ``Sequential\; Colormap''}$\}$.
The user would like to get heatmaps with the same form as the query chart but with more contrast colors. 
S/he can get the desired results by changing \textit{Type} and \textit{Color} attributes to $\{$\emph{Type: ``Heatmap'', Color: ``Diverging\; Colormap''}$\}$ in the retrieval stage.

\subsubsection{Fuzzy Retrieval by User Intent}

Designing an expressive visualization incurs a steep learning rate for some novices, who may not have a specific design prior and prefer to perform trial and error on the search engine to seek a desirable design.
Below, we list three scenarios to illustrate how our method supports such retrieval.
Similarly, we will compare the results of the intent-free and intent-aware retrieval.
\begin{figure}[t]
    \centering
    \includegraphics[width=0.98\linewidth]{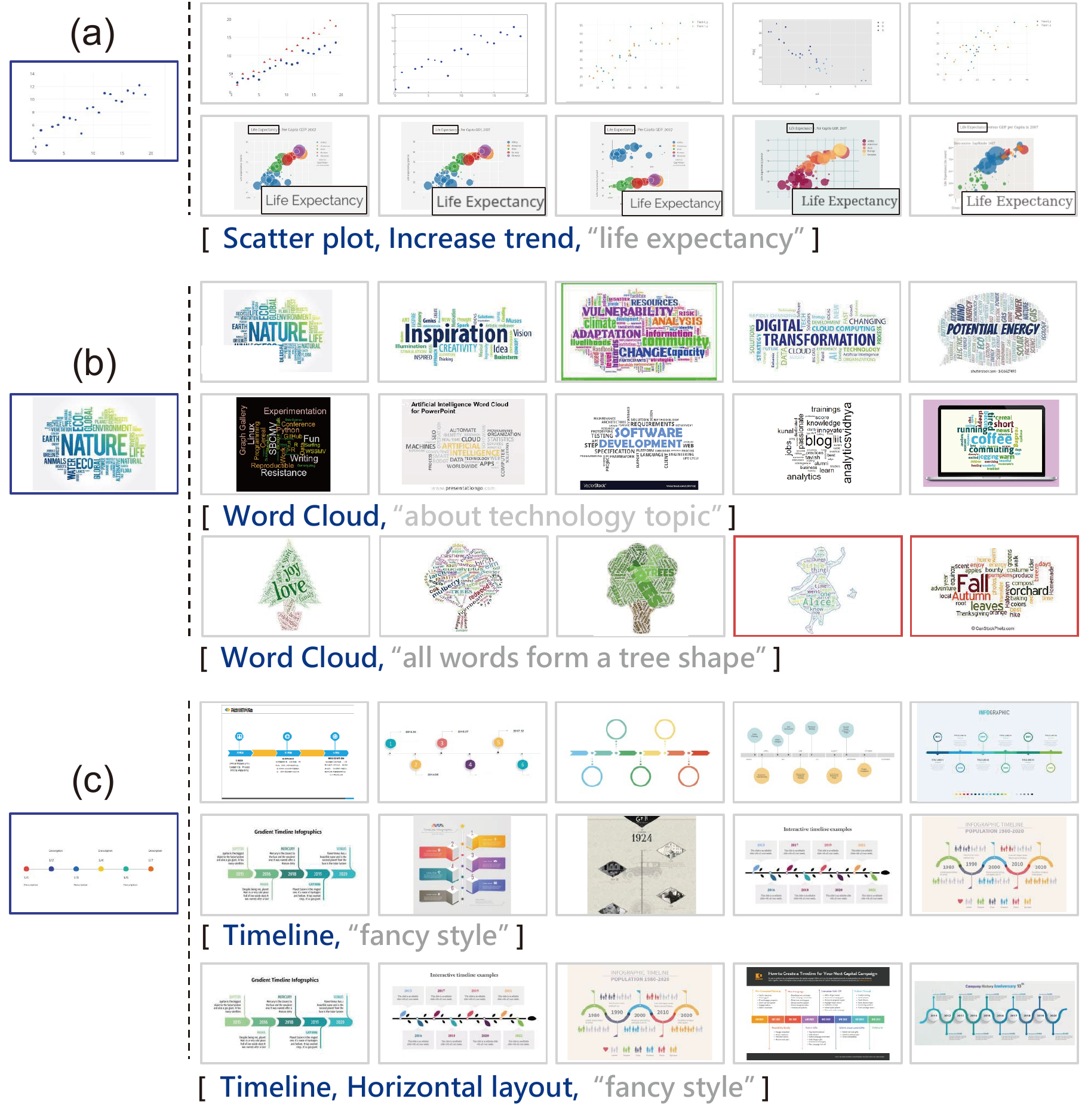}
    \vspace{-2mm}
    \caption{\textbf{Three cases for showing fuzzy retrieval.}
    \revise{Negative examples are highlighted with a red border.}}
    \label{fig:case2}
    \vspace{-5mm}
\end{figure}

\noindent \textbf{Text Information Seeking.}
Keyword-based searching is popular, but existing works approach this goal by storing text information in advance~\cite{hoque2019searching, chen2015diagramflyer} or relying on OCR \cite{savva2011revision}, hindering generalizability and robustness of the works.
Instead, our method allows to recognize text information based on knowledge transferred by the pretrained CLIP model.
As shown in Figure~\ref{fig:case2} (a), to generate a chart for visualizing \textit{``life expectancy''}, the user keeps the origin chart type \textit{Scatter plot} and data trend \textit{Increase trend}, and adds the user intent as a text prompt {\prompt{``life expectancy''}}.
\revise{The intent-aware retrieval results imitate the famous design in Gapminder.}

\noindent \textbf{Relevant Topic Finding.}
Our method also has the capacity to find chart examples related to a specific topic.
Previous work used word2vec~\cite{church2017word2vec} to model the distances between words~\cite{hoque2019searching}, limiting the search scope within texts.
We enhance the generalizability of matching text prompt with both text information and visual attributes.
In Figure~\ref{fig:case2} (b), given a word cloud with the theme of nature, the user can retrieve some examples with text prompt as {\prompt{``about technology topic''}}.
Our system returns examples with content about technology.
Moreover, with the text prompt of {\prompt{``all words form a shape of a tree''}}, our system aligns the semantic information of text with visual attributes, returning examples with tree shapes instead of only containing the word ``tree''.

\noindent \textbf{Abstract Description Searching.}
In the process of retrieval, users not having a clear search target tend to use ambiguous and abstract descriptions.
In Figure~\ref{fig:case2} (c), the user inputs a basic timeline diagram with {\prompt{``fancy style''}}, and gets several well-designed examples.
To constrain the scope of \textit{vertical layout}, the user adds such attribute and performs targeted retrieval.
\subsection{Qualitative Evaluation}
\label{ssec:Qualitative}
\subsubsection{Study Design}
Our study recruited \revise{seven participants (three women, four men)} with ages 22 to 26. To demonstrate that our framework is useful for different levels of users, we enroll four experts, \textit{i.e.}, visualization designers (E1 - E4), and three novices with only experience in using visualizations in reports or courses (N1 - N3). 

Before the experiment, we collected basic information from the participants through a questionnaire and introduced them to the use scenarios of chart retrieval.
Then we instructed them to familiarize themselves with the system workflow.
We also showed some examples of using prompts to help participants understand the prompt better and make a better choice of prompts.
After understanding the background, participants were allowed to freely explore and use our system. 
They could upload a query chart and try to get a satisfactory design in multiple iterations. 
If there was no satisfactory design, they could also try combinations of attributes and prompts to see if the retrieved chart has inspiring results.
Throughout the process, participants were encouraged to think aloud and give feedback whenever they wanted.
After they felt it was sufficient, we conducted interviews about the usability of our system. 
Three questions were included: 1) Whether our approach is \emph{effective} in helping them find the chart that meets their intents; 2) Whether our approach is \emph{efficient} enough to avoid long waits; 3) Whether they are \emph{satisfied} with our system.
The experiment lasted about 45 minutes. \revise{The participants were not compensated with money but beverages worthing about \$5 dollars after the study}.

\subsubsection{Feedback}

\noindent \textbf{Effectiveness.}
In general, all users agreed that our system effectively supported for retrieving the desired charts based on their intents. 
Most participants (6 out of 7) thought that our setting of the primary attribute was appropriate and comprehensive. E2 suggested to extract the main hue used in the chart in the annotation stage.
Regarding the comprehensiveness and accuracy of the prompts, all participants felt that their search needs were largely met and the results were generally consistent with their intents.
N1 spent a long time exploring the map \revise{as shown in Figure~\ref{fig:case1} (a)}. He found it exciting when typing ``India'', an India map was retrieved. 
Our system also effectively recognized his intents when he searched for maps of various colors.

The participants pointed out some limitations as well.
When the prompt given by the user had multiple meanings, sometimes only one or two of the search results were exactly what the user expects. 
E4 tried to add a trend arrow on the bar chart. 
However, when he only typed ``with arrow'', some bar charts using arrow icons were also retrieved.
He said that sometimes it was required to go and try several prompts to get the desired result.
E3 encountered a similar situation, and it might be due to our dataset only containing limited charts that matched her specifications.
Besides, sometimes participants may want to adjust attributes in the query chart, such as changing the white background to dark (N2), and changing the colorful design to black and white (E2). 
However, one or two of the returned results still had the features that they would like to change.
``There are times when I want the query to focus more on the prompt I give and less on the original chart,'' said N2.

\noindent \textbf{Efficiency and Satisfaction.}
All participants agreed that our system was responsive and user friendly.
They were satisfied with our retrieving framework, and they pointed out some possible improvements.
They found it helpful to decouple a chart into attributes and have a selective query.
N2 stated, ``It saved me the time of checking a lot of examples and found the desired design directly''.
N3 appreciated the function of adding a new classifier since it can be used to explore new types of visualizations. ``It is convenient for novices who do not understand the type and content of the chart at first,'' he gave the reason.
E1, a participant with a design background, said our results were inspiring.
She was willing to see more results, even if they did not exactly match the search intents.
``This could inspire me to come up with new ideas,'' she added.
\section{Discussion}
WYTIWYR is an intent-aware chart query framework that
can address various issues related to traditional chart retrieval and expedite the design process.
In conventional retrieval, not all chart attributes may be relevant to users' intents, whilst some preferred attributes may even be absent from the chart.
Instead, our system disentangles and combines attributes to ensure that the retrieved results encompass all attributes of user's intents.
There are also several limitations in our current methods and avenues for future work.

\subsection{Limitations}

\noindent
\textbf{User Intent or Query Chart.}
The input of our method consists of a query chart and user intent. The composition of two factors affects the retrieval results.
As shown in Figure~\ref{fig:case1} (b), users have a strong intent of charts that have a curve, and our system would tend to return the results meeting such intents. However, the perception of global distribution may be lost.
The priority between \revise{user intents as text and global distribution in the input chart} is hard to tackle due to the usage scenarios.
Furthermore, within the user intent, the weights of selected attributes and text prompts are hard to set in the similarity modeling.
A dynamic adjustment of the priority between the user intent and query chart is demanded in this scenario.

\begin{figure}[t]
    \centering
    \includegraphics[width=0.98\linewidth]{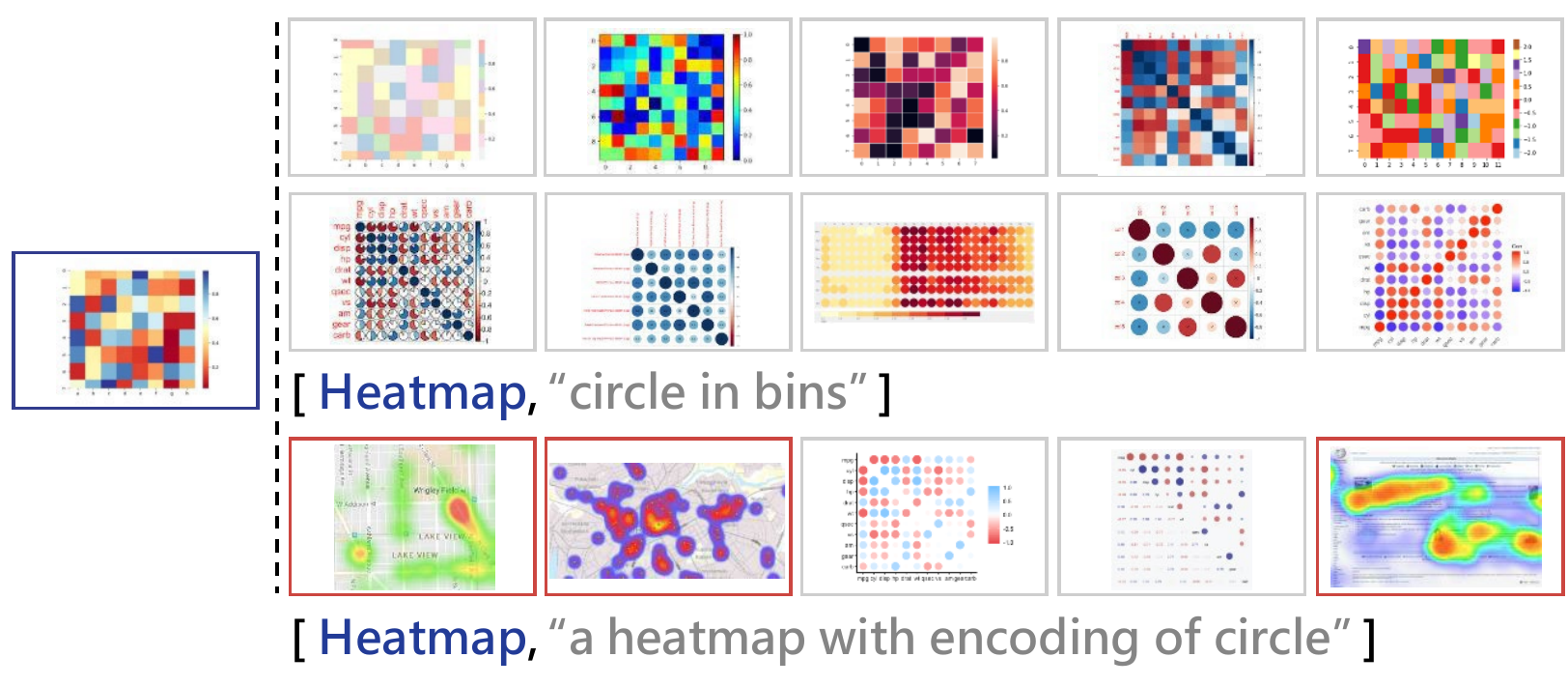}
    \vspace{-2mm}
    \caption{\textbf{A case for revealing the limitations.} \revise{Negative examples are highlighted with a red border.}}
    \label{fig:case-limit}
    \vspace{-4mm}
\end{figure}

\noindent
\textbf{Prompt Sensibility.}
Text prompt is integrated with query chart as joint input for retrieval in our work.
However, designing effective prompts is challenging due to the significant impact of even slight wording changes on performance\cite{zhou2022learning}.
In our framework, prompt design is limited by the CLIP pretrained model, which is trained mainly on natural images, which makes it difficult to accurately interpret expert expressions for visualization domain.
For instance, Figure~\ref{fig:case-limit} shows that using a more professional prompt in the third row results in negative results, when compared to the second row with identical intent.
\revise{The CLIP model also struggles to identify rare or novel objects. Approaches for fine-tuning~\cite{li2022blip, li2021align, zhang2022tip} the CLIP model or using advanced language-image models are promising.
Obtaining enough chart-description pairs for training is also necessary, but the process is time-consuming and resource-intensive.}

\noindent
\textbf{Dataset.}
Despite collecting a large number of real-world charts, our database remains limited in meeting vast user needs. \revise{In Figure~\ref{fig:case2} (b), only the first three charts are tree-shaped since we only have these charts that meet the user's intent in our database}. The unbalanced dataset may adversely affect retrieval performance, as the attribute classifier training could become biased towards classes with larger volumes.
We alleviate it by using focal loss, which assigns higher weights to minority class and misclassified examples.
Due to the lack of a more detailed category in
our dataset, some inaccurate results will occur in the retrieval.
In Figure~\ref{fig:case-limit}, two types of heatmaps are visible in the retrieval results.

\noindent
\revise{\textbf{Chart Attributes.}
In the preliminary study, we employed several chart types to determine the chart attributes that users tend to concentrate on. 
We randomly chose samples from both synthesized datasets and real-world collections to diversify the chart attributes.
Nevertheless, the attributes may not fully cover user intents. A more comprehensive survey will be conducted to address the issue.}

\subsection{Future Work}

\noindent
\textbf{Trade-off Control.}
To balance between user intent and query chart, we plan to add a slider equipped with dynamic weight control to let the user manage this trade-off by themselves in the near future.

\noindent
\textbf{Text Prompt Auto-completion.}
Previous works~\cite{setlur2020sneak, wang2022towards} utilize auto-completion as a hint to assist users in the application process, which is also helpful for our system to alleviate the interpretability shift between the user side and the model side.
Specifically, we aim to build a mapping table to store the relationship between appropriate prompts and common user text expressions.
Then, when a user completes the input, the keywords of it will be extracted by the named entity recognition technique~\cite{nadeau2007survey}.
Finally, the corresponding optimal prompt would be automatically completed by the search in the mapping table.

\noindent
\textbf{Text-only retrieval.}
Our system currently supports chart-only retrieval and chart-text retrieval.
To offer a more generalized retrieval framework, we plan to add a text-only query, allowing users to obtain their retrieval by only providing text input.
The text-only retrieval is beneficial to users who do not have any reference chart.
To this end, we aim to prepare several basic charts with primary attributes in advance to serve as temporary query charts, which can be replaced by a more reliable chart from returned results by retrieval. The process can be regarded as an iteration of desirable results, with a clearer search goal and narrowed search scope.

\section{Conclusion}
In this paper, we propose a user intent-aware chart retrieval framework, which leverages multi-modal input to fuse explicit visual attributes and implicit user intent into the retrieval process.
This pipeline consists of two core stages, namely \textit{Annotation}, and \textit{Retrieval}.
The \textit{Annotation} stage disentangles visual attributes in the query chart to ensure a flexible combination of user intent attributes used in retrieval.
The \textit{Retrieval} stage allows users to integrate the text prompt with the query chart to achieve more customized retrieval.
Quantitative experiments prove the superior performance of our method compared with previous methods.
Furthermore, we conduct two case studies containing two common retrieval strategies and interviews to demonstrate the effectiveness and usability.
As an initial step to fuse user prompt in chart retrieval, we hope to enhance the prompt capacity to better meets users' growing needs and various usage scenarios.
\revise{Dataset, code, pretrained model have been released to promote future research in this direction.}

\section{Acknowledgments}
The authors wish to thank anonymous reviewers for their constructive comments. 
The work was supported in part by National Natural Science Foundation of China (62172398), and the Red Bird Program at the Hong Kong University of Science and Technology (Guanghzou).

\bibliographystyle{eg-alpha-doi}
\bibliography{EGauthorGuidelines-eurovis23-full}
\end{document}